\pdfoutput=1
\documentclass{article}
\usepackage[T1]{fontenc}
\usepackage{spconf,amsmath,graphicx}
\usepackage{booktabs}
\usepackage{url}
\newcommand{\rtfslowfull}{2.09$\times$}
\newcommand{\rtfslownovoc}{1.47$\times$}
\newcommand{\energycost}{43\%}
\newcommand{\stfpwer}{0.028}
\newcommand{\stdynwer}{1.11}
\newcommand{\stnovocutmos}{\mbox{-0.041}}
\newcommand{\dynmmutmos}{4.467}
\newcommand{\dynmmwer}{0.029}
\newcommand{\kofpcer}{0.052}
\newcommand{\kodyncer}{1.27}
\newcommand{\stfputmos}{4.473}
\newcommand{\stmixrecover}{4.252}
\newcommand{\ovfputmos}{4.398}
\newcommand{\kkfputmos}{4.524}
\newcommand{\ovgranpt}{1.363}
\newcommand{\ovgrang}{3.761}
\newcommand{\kkptwerfp}{0.030}
\newcommand{\kkptwer}{0.261}
\newcommand{\stpteightutmos}{\mbox{-1.642}}
\newcommand{\stwsixutmos}{\mbox{-0.56}}
\newcommand{\ovfpwer}{0.032}
\newcommand{\stwfourcost}{2.8}
\newcommand{\ovgptqdelta}{\mbox{-0.029}}
\newcommand{\ovawqdelta}{\mbox{-0.100}}
\newcommand{\ovrtnglmdelta}{\mbox{-0.093}}
\newcommand{\ovrtnpclmdelta}{\mbox{-0.28}}
\newcommand{\ovheadplain}{\mbox{-0.99}}
\newcommand{\ovheadsmooth}{\mbox{-0.64}}
\newcommand{\ovheadgone}{\mbox{-0.39}}
\newcommand{\seedsdmax}{0.06}
\newcommand{\nsent}{200}
\newcommand{\nboot}{10{,}000}
\newcommand{\ovprojdelta}{\mbox{-0.17}}
\newcommand{\ovcodecdelta}{\mbox{-0.68}}
\newcommand{\ovcodecgdelta}{\mbox{-0.12}}
\newcommand{\kkptdelta}{\mbox{-3.15}}
\newcommand{\actptvoc}{\mbox{-2.98}}
\newcommand{\actptvocwer}{1.30}
\newcommand{\actpcvoc}{\mbox{-0.20}}
\newcommand{\acttokvoc}{\mbox{-0.08}}
\newcommand{\ovactptcodec}{\mbox{-0.14}}
\newcommand{\ovactptlm}{\mbox{-0.95}}
\newcommand{\sttwowfour}{\mbox{-0.64}}
\newcommand{\sttwodec}{\mbox{-0.11}}
\newcommand{\sttwonodec}{\mbox{-0.20}}
\newcommand{\kydep}{\mbox{-2.99}}
\newcommand{\kydepsix}{\mbox{-1.38}}
\newcommand{\listenraters}{12}
\newcommand{\listenclips}{96}
\newcommand{\listenrho}{0.91}
\newcommand{\orpintlat}{1.5$\times$}
\newcommand{\orpintenergy}{34\%}
\newcommand{\crestpcgain}{10.6}
\newcommand{\actbestkkrecipe}{\mbox{-0.04}}
\newcommand{\actbestkkrecipea}{\mbox{-0.13}}
\newcommand{\actbestovrecipe}{\mbox{-0.11}}
\newcommand{\actbestovrecipea}{\mbox{-0.18}}
\newcommand{\actbestwerbound}{0.01}
\newcommand{\actptovcodec}{\ovactptcodec{}}
\newcommand{\actptovlm}{\ovactptlm{}}
\newcommand{\kyfpwer}{0.034}
\newcommand{\kydepeight}{\mbox{-0.02}}
\newcommand{\kydepgthirtytwo}{\mbox{-2.99}}
\newcommand{\ovbfmatchedlat}{0.96$\times$}
\newcommand{\ovbfmatchedenergy}{0.98$\times$}
\newcommand{\ovmatchedlat}{1.11$\times$}
\newcommand{\ovmatchedenergy}{1.27$\times$}
\newcommand{\kkwfourcost}{0.07}
\newcommand{\wfoureightmax}{0.09}
\newcommand{\actptovproj}{\mbox{-2.31}}
\newcommand{\cihwst}{0.03}
\newcommand{\cihwkk}{0.01}
\newcommand{\cihwov}{0.09}
\newcommand{\cihwrep}{0.16}
\newcommand{\ovnocodecdelta}{\mbox{-0.93}}
\newcommand{\cblm}{\mbox{-0.01}}
\newcommand{\cbflow}{\mbox{-2.17}}
\newcommand{\cbflowg}{\mbox{-0.55}}
\newcommand{\cbrecipe}{\mbox{-0.02}}
\newcommand{\cbintfour}{0.00}
\newcommand{\vxfpwer}{0.043}
\newcommand{\vxdit}{\mbox{-2.73}}
\newcommand{\vxditwer}{0.14}
\newcommand{\vxvae}{\mbox{-1.25}}
\newcommand{\vxditg}{\mbox{-1.26}}
\newcommand{\vxvaeg}{\mbox{-0.25}}
\newcommand{\vxrecipeg}{\mbox{-0.21}}
\newcommand{\vxintfour}{0.03}
\newcommand{\fvocplaing}{\mbox{-0.48}}
\newcommand{\fvocgptqg}{\mbox{-0.07}}
\newcommand{\kydepsmoothwer}{0.13}
\newcommand{\kydepplaing}{\mbox{-2.98}}
\newcommand{\kydepgptqg}{\mbox{-0.08}}
\newcommand{\kydepplainwer}{1.14}
\newcommand{\ncalib}{64}
\newcommand{\aeightpcmax}{0.17}
\newcommand{\aeightptf}{\mbox{-3.10}}
\newcommand{\aeightptky}{\mbox{-2.99}}
\newcommand{\aeightptcsm}{\mbox{-2.64}}
\newcommand{\aeightptzn}{\mbox{-1.36}}
\newcommand{\aeightptrestmin}{0.08}
\newcommand{\aeightptrestmax}{0.29}
\newcommand{\aeightrecipemax}{0.12}
\newcommand{\stnbitsall}{\mbox{-2.38}}
\newcommand{\stnbitsnovoc}{\mbox{-0.07}}
\newcommand{\stnbitsnovocwer}{0.033}
\newcommand{\stsimnovocg}{\mbox{-0.10}}
\newcommand{\stnbitscover}{99.5\%}
\newcommand{\ovintfour}{\mbox{-0.14}}
\newcommand{\ovsimlmg}{\mbox{-0.09}}
\newcommand{\ovintfourkocer}{+0.013}
\newcommand{\ovintfourkocerci}{[+0.001, +0.029]}
\newcommand{\orintfour}{0.01}
\newcommand{\kyintfour}{\mbox{-2.81}}
\newcommand{\csmintfour}{\mbox{-1.02}}
\newcommand{\ovintfourlat}{2.1$\times$}
\newcommand{\ovintfourenergy}{2.2$\times$}
\newcommand{\orintfourlat}{0.50$\times$}
\newcommand{\orintfourenergy}{0.45$\times$}
\newcommand{\nisqarhomin}{0.79}
\newcommand{\nisqarhomax}{0.97}
\newcommand{\nisqarhosttwo}{0.46}
\newcommand{\nproxyruns}{251}
\newcommand{\nseedcond}{22}
\newcommand{\nseedmodels}{ten}
\newcommand{\stnbitsrtfratio}{0.60$\times$}
\newcommand{\stnbitsrsssave}{41\%}
\newcommand{\vxditgptqg}{\mbox{-0.41}}
\newcommand{\vxallcalib}{\mbox{-1.01}}
\newcommand{\cbrest}{\mbox{-0.42}}
\newcommand{\vxrest}{\mbox{-1.75}}
\newcommand{\stvocdelta}{\mbox{-2.87}}
\newcommand{\stflowdelta}{\mbox{-0.17}}
\newcommand{\aeightweightst}{\mbox{-0.01}}
\newcommand{\aeightweightkk}{\mbox{-0.07}}
\newcommand{\aeightweightov}{\mbox{-0.06}}
\newcommand{\actbeststrecipe}{\mbox{-0.10}}
\newcommand{\actbeststrecipea}{\mbox{-0.10}}
\newcommand{\vxminormax}{0.10}
\newcommand{\ffpwer}{0.035}
\newcommand{\kkwsix}{\mbox{-0.003}}
\newcommand{\ovwsix}{\mbox{-0.03}}
\newcommand{\znrest}{\mbox{-0.05}}
\newcommand{\diarest}{\mbox{-0.27}}

\title{Same Bit Width, Different Outcomes:\\
Post-Training Quantization of Text-to-Speech Across Architectures}

\name{Se Un Park$^{\ast}$, Yutae Kim, Junyoung Park$^{\ast}$\thanks{$^{\ast}$Corresponding authors.}}
\address{UX Factory, Inc.}

\begin{document}
\ninept
\setlength{\textfloatsep}{12pt plus 2pt minus 4pt}
\setlength{\abovecaptionskip}{4pt}
\maketitle

\begin{abstract}
Post-training quantization (PTQ) reduces the cost of on-device
text-to-speech (TTS), but published evaluations cover one system or method.
We evaluate PTQ across TTS architectures under one protocol with three
core models, weight and activation ablations of eight more, and two
held-out models quantized blind. Four-bit per-channel
weights reduce UTMOS, a predicted mean opinion score, by \stwfourcost{} on
Supertonic and \kkwfourcost{} on Kokoro, and per-tensor scaling can
cause severe degradation even at 8 bits. The same bit width yields different outcomes, because the
sensitive component is model-specific and not reliably predicted from
the model class. A staged ablation procedure identifies it, and
per-layer GPTQ can restore it to within 0.1 UTMOS. Real int8 and int4 kernels reproduce
the simulated ordering at hardware-dependent cost. On a Mac mini, a
4-bit weight kernel runs Supertonic at \stnbitsrtfratio{} the fp32 latency while int8 is slower, so each configuration requires
validation on the target runtime.
\end{abstract}

\begin{keywords}
speech synthesis, post-training quantization, on-device inference, model
compression, edge computing
\end{keywords}

\section{Introduction}
\label{sec:intro}

On-device \mbox{text-to-speech} (TTS) enables private and offline speech
synthesis under memory and compute constraints~\cite{achanta2021ondevice}.
Post-training quantization (PTQ) compresses a trained model without
retraining~\cite{frantar2023gptq,lin2024awq}, but the bit width is only
one of several design choices. Scale sharing, the quantized components, the activation precision, and the operator coverage of the
runtime each affect the outcome. Prior quantization studies cover a single
system, or several systems under a single
method~\cite{kawamura2025bittts,vora2024ptq4adm,khandelwal2025adit}, and
no study compares heterogeneous pretrained TTS pipelines,
quantized-component scopes, activation granularity, and measured
deployment paths under one protocol.

This paper provides that comparison and shows that the same bit width
yields different outcomes because the sensitive component is
model-specific. The study covers three core models, eight replication models, and two models that were held out and
quantized blind.
\begin{itemize}
\item \textbf{A cross-architecture sensitivity map.} Component ablations
  under one protocol reveal model-specific weight sensitivity and
  interactions that amplify the whole-model degradation. Mixed-precision configurations reduce the UTMOS degradation, and
  per-layer GPTQ can restore the sensitive component.
\item \textbf{Activation and combination results.} Activation sensitivity
  depends on the scale granularity and on the quantized components.
  Per-channel 8-bit activations add at most \wfoureightmax{} UTMOS loss
  to the selected 4-bit configurations of the core models.
\item \textbf{Deployment measurements.} Real int8 and int4 execution
  shows that peak memory, latency, and energy depend on the runtime and
  the hardware (Sec.~\ref{ssec:deployment}), which motivates the staged procedure tested blind in
  Sec.~\ref{ssec:blind}.
\end{itemize}

\section{Related work}
\label{sec:related}

TTS compression includes architecture search in
LightSpeech~\cite{luo2021lightspeech}, acoustic-model int8 PTQ combined with architectural
compression~\cite{jain2025compact}, and BitTTS~\cite{kawamura2025bittts},
which applies extreme ternary quantization (1.58-bit training) and
weight indexing to build compact TTS models for on-device use and
identifies the vocoder as the sensitive component of its model. PTQ for diffusion models is
established for image
generation~\cite{shang2023ptq4dm,li2023qdiffusion,feng2024mpqdm} and has
been extended to audio generation by PTQ4ADM and timestep-aware
quantization~\cite{vora2024ptq4adm,khandelwal2025adit}. HAWQ ranks layers by Hessian curvature for mixed precision in
vision~\cite{dong2019hawq}, and Diet-KIT selects per-layer precision by
sensitivity for speech translation~\cite{liu2026dietkit}.

\section{Experimental setup}
\label{sec:setup}

\begin{figure}[t]
  \centering
  \includegraphics{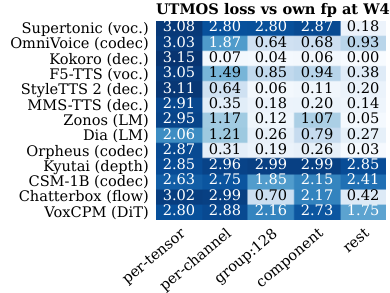}
  \caption{Sensitivity map. Magnitude of the paired UTMOS loss at 4-bit
  weights against each model's own fp baseline for the whole model under
    three scale granularities, for the most sensitive component alone at
  W4 per-channel (component, named in the row label), and for the rest
    of the model at W4 per-channel with that component at fp (rest).
  OmniVoice uses its CUDA fp16 baseline, and the last two rows are the
  held-out models of Sec.~\ref{ssec:blind}.}
  \label{fig:map}
\end{figure}

\textbf{Models and data.}
The core systems are Supertonic V3 (99M)~\cite{supertone2026supertonic},
a flow-matching model with a vocoder, OmniVoice
(0.6B)~\cite{zhu2026omnivoice}, a masked-diffusion language model (LM)
with a token head and a codec decoder, and the feedforward model Kokoro
(82M)~\cite{hexgrad2025kokoro}. Their English floating-point (fp) baselines reach UTMOS \stfputmos{},
\ovfputmos{}, and \kkfputmos{} at WER \stfpwer{}, \ovfpwer{}, and
\kkptwerfp{}, respectively, on ONNX CPU fp32, torch Metal fp16, and torch
x86 fp32, with a default number of function evaluations (NFE) of 8 for
Supertonic and 32 for OmniVoice. The replication models
(Fig.~\ref{fig:map}) are the flow-matching F5-TTS~\cite{chen2024f5tts},
the feedforward StyleTTS~2~\cite{li2023styletts2} and
MMS-TTS~\cite{pratap2024mms}, and the autoregressive codec-token LMs
Zonos~\cite{zyphra2025zonos}, Dia~\cite{narilabs2025dia},
Orpheus~\cite{canopy2025orpheus}, Kyutai TTS~\cite{kyutai2025tts}, and
CSM-1B~\cite{sesame2025csm}. Two further models were held out until the
procedure of Sec.~\ref{ssec:blind} was fixed and then evaluated blind. Chatterbox~\cite{chatterboxtts2025} combines a Llama backbone
(T3, 503M) that emits speech tokens with a flow-matching decoder (S3Gen,
112M) and a HiFT vocoder (21M). In VoxCPM-0.5B~\cite{zhou2025voxcpm},
a MiniCPM-4 LM (434M) and a residual acoustic LM (90M) drive a local
encoder (60M), a local diffusion transformer (DiT, 64M) that emits
patches of continuous latents, and a convolutional audio VAE decoder
(75M). Each evaluated language uses the same \nsent{} frozen FLORES-200
devtest sentences~\cite{nllb2022flores} with fixed per-utterance seeds,
in English and, where supported, in Korean. Each condition is paired with the unquantized model at its native
floating-point precision (fp)\footnote{fp32 for Supertonic, Kokoro,
F5-TTS, StyleTTS~2, MMS-TTS, and Chatterbox, fp16 for OmniVoice and Dia,
and bf16 for Orpheus, Kyutai, CSM, VoxCPM, and the Zonos backbone, as
shipped.} on the same platform and at the same number of sampling
steps.\footnote{Calibration uses FLORES-200 dev
sentences that are disjoint from the evaluation set. OmniVoice, which is
evaluated in both languages, uses 64 English and 64 Korean sentences, and
the English-only models of Secs.~\ref{ssec:calibrated}
and~\ref{ssec:blind} use 64 English sentences. Code, sentences, per-utterance scores, timing records, pinned versions,
seeds, calibration settings, and the quantization specification of every
run are available at
\url{https://github.com/uxfacdev/tts-ptq-map}.}

\textbf{Quantization.}
The weight-only sweep applies symmetric round-to-nearest (RTN)
quantization at 8, 6, and 4 bits with $s=\max|w|/(2^{b-1}-1)$ and 16-bit
scale factors shared per tensor, per output channel (the default), or per
group of 128 consecutive weights within a channel (group:128). Channels
denote feature dimensions rather than audio channels. A weight scale is shared per output feature of a linear layer or per
output filter of a convolution. A per-channel activation scale is shared
per input feature over all time steps, and a per-token activation scale
is shared per time step. Linear
and convolution weights are quantized, whereas biases, embeddings, and
normalization parameters remain in floating point. We quantize the whole
model, individual components, and mixed-precision combinations, and we sweep the NFE over $\{4,8,12\}$ for Supertonic and $\{4,8,16,32\}$ for OmniVoice. Simulated conditions dequantize the weights before execution and
measure quality only. Calibrated PTQ applies per-layer GPTQ~\cite{frantar2023gptq} or
activation-aware weight scaling~\cite{lin2024awq,xiao2023smoothquant} to
one component at a time, with input statistics collected from \ncalib{}
English calibration sentences synthesized by the fp model
(Sec.~\ref{ssec:calibrated}). Every other module of that component is quantized by RTN at the same
bit width and granularity. Simulated
dynamic 8-bit activations (A8) are added to W8 per-channel weights and to
the W4 group:128 configurations. For Supertonic the vocoder is excluded from weights and activations,
for Kokoro every module is quantized, and for OmniVoice the W4
configuration (LM at W4, token head at W8, codec at fp) adds A8 to every
module, on x86 fp32 baselines for Supertonic and Kokoro and on a CUDA
fp16 baseline for OmniVoice. The replication models receive whole-model A8 at per-channel and
per-tensor scales and A8 on their W4 configurations. Real
4-bit weight execution uses the block-wise 4-bit operator of ONNX Runtime
(\texttt{MatMulNBits}, block size 128) for Supertonic on x86, after the
dense $1\times1$ convolutions of its flow estimator are rewritten as
matrix products so that the kernel covers \stnbitscover{} of the weights
(7.6\% as shipped), and torchao int4 weight-only quantization
(group 128, bf16 operands) for the LMs of OmniVoice, Orpheus, Kyutai,
CSM, Chatterbox, and VoxCPM on an RTX PRO 6000.

\textbf{Quality and system measurements.}
UTMOS~\cite{saeki2022utmos}, a predicted mean opinion score, is the
proxy for English naturalness.
NISQA-TTS~\cite{mittag2020nisqatts} and DNSMOS
P.835~\cite{reddy2022dnsmos} on \nproxyruns{} runs are secondary
proxies, compared with UTMOS by the Spearman correlation of paired deltas
over the conditions of one model. The transcription
error of faster-whisper large-v3~\cite{radford2023whisper} serves as the
intelligibility proxy, reported as the per-utterance mean word error rate
(WER) for English and character error rate (CER) for
Korean.
Paired 95\% bootstrap intervals use \nboot{} resamples over the
\nsent{} sentences, conditional on the evaluated settings. A quantized
configuration is defined to be quality-preserving on the two English
proxies when its paired $\Delta$UTMOS interval lies within $[-0.05,
0.05]$ and the upper endpoint of its paired $\Delta$WER interval is at
most $0.01$ relative to fp, evaluated at unrounded endpoints.

Real CPU execution uses ONNX Runtime dynamic int8 for Supertonic and
PyTorch dynamic int8 for Kokoro on a Mac mini M4 Pro (4 threads, 20 sentences, 5 repeats). GPU measurements use torchao 0.18 with compiled execution on
A100, L4, and RTX PRO 6000 devices. The LM of OmniVoice runs at 8-bit
weights and activations (W8A8) and the LM of Orpheus at weight-only
int8, with the codecs in floating point, bf16 operands, and compiled
fp16 (OmniVoice) or bf16 (Orpheus) latency baselines. We measure latency as the real-time factor (RTF), the peak CPU resident
set size (RSS), and the energy per second of audio. On the CPU, the
energy is the whole-chip energy above idle over the complete 20-sentence
process including model loading, whereas the RTF counts synthesis time
only, so $E \neq P \times \mathrm{RTF}$. On the GPU, the energy is the
NVML device energy without host power over twelve sentences and three
repeats (twenty and five at the matched operating point), excluding
loading and compilation.

\section{Results}
\label{sec:results}

\subsection{Scale granularity and bit width}
\label{ssec:granularity}

8-bit per-channel weights produce small changes on all three core
models, and the paired 95\% intervals include zero. 6-bit weights yield \stwsixutmos{} $\Delta$UTMOS on Supertonic, \kkwsix{}
on Kokoro, and \ovwsix{} on OmniVoice. 4-bit weights reduce UTMOS
by \stwfourcost{} on Supertonic and by \kkwfourcost{} on Kokoro, whereas
OmniVoice degrades substantially (Fig.~\ref{fig:map}). At 4 bits,
group:128 scaling raises the UTMOS of OmniVoice from \ovgranpt{} under
per-tensor scaling to \ovgrang{} but does not recover Supertonic. A
weight-domain signal-to-quantization-noise ratio estimate, in which the
largest weight of a scale-sharing group sets the quantization step,
gives per-channel scales a \crestpcgain{}~dB advantage over per-tensor
scales on two released checkpoints, more than one additional bit
provides. Per-tensor scaling also degrades Kokoro (\kkptdelta{}
$\Delta$UTMOS, WER \kkptwerfp{} to \kkptwer{}) and W8 Supertonic
(\stpteightutmos{} at fp-level WER).

\subsection{Component sensitivity and mixed precision}
\label{ssec:component}

The vocoder of Supertonic alone yields \stvocdelta{} $\Delta$UTMOS at
W4, whereas its flow estimator alone yields \stflowdelta{}. Keeping the vocoder at W8 with the rest at W4 restores UTMOS to \stmixrecover{} against \stfputmos{}
at fp. The CUDA ablation of OmniVoice attributes more of the degradation
to the codec decoder (\ovcodecdelta{}) than to the token head
(\ovprojdelta{}) or the LM (\ovrtnpclmdelta{}). Two components quantized together degrade UTMOS by more than the sum of
their individual losses, with \ovheadplain{} for the token head with the
codec and \ovnocodecdelta{} for the LM with the token head (the rest
column of Fig.~\ref{fig:map}). Group:128 scaling improves the decoder-only result to
\ovcodecgdelta{}, whereas the vocoder of Supertonic remains degraded. The paired 95\% intervals lie within $\pm$\cihwst{},
$\pm$\cihwov{}, and $\pm$\cihwkk{} UTMOS of their estimates for
Supertonic, OmniVoice, and Kokoro, respectively, and within
$\pm$\cihwrep{} for the replication models.

Across the replication models (Fig.~\ref{fig:map}), group:128 scaling
reduces the degradation of Zonos, StyleTTS~2, and F5-TTS without
establishing equivalence to fp, and it has no effect on Kyutai. For
Zonos and Dia the LM carries the loss, since the rest of the model with
the LM at fp loses only \znrest{} and \diarest{} (Fig.~\ref{fig:map}).
Group:32 scaling does not improve the depth
transformer of Kyutai (\kydepgthirtytwo{} against \kydep{} per-channel), whereas 6-bit and
8-bit weights yield \kydepsix{} and \kydepeight{}.
The W4 decoder of StyleTTS~2
and the rest of its network yield \sttwodec{} and \sttwonodec{} separately
but \sttwowfour{} together.

\subsection{Sampling steps and matched baselines}
\label{ssec:steps}

\begin{figure}[t]
  \centering
  \includegraphics{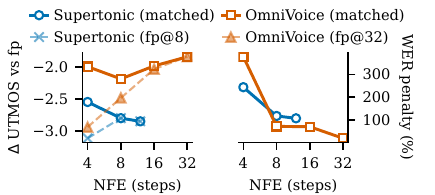}
    \caption{W4 per-channel versus sampling steps (NFE). Left, $\Delta$UTMOS of the W4 run against fp at the same
  NFE (matched, solid) and against the single default-step fp run, fp@8
  for Supertonic and fp@32 for OmniVoice (dashed). Right, the matched
    relative WER penalty in percent, the mean per-utterance WER of W4 over
  that of fp at the same NFE, minus one.}
  \label{fig:steps}
\end{figure}

Comparing every step count against a single default-step fp baseline
makes the quantization degradation of OmniVoice appear to decrease with
the number of steps (Fig.~\ref{fig:steps}), and pairing each quantized
run with fp at the same NFE removes this confound.
Additional steps reduce the WER penalty under W4 (right panel) without
closing the UTMOS gap (left panel), which widens with the number of steps
on Supertonic.

\subsection{Calibration controls}
\label{ssec:calibrated}

On the LM of OmniVoice, W4 group:128 GPTQ~\cite{frantar2023gptq} yields
\ovgptqdelta{} $\Delta$UTMOS against \ovrtnglmdelta{} for RTN, and
activation-aware scaling~\cite{lin2024awq,xiao2023smoothquant} yields
\ovawqdelta{}. On the token head and codec together, the best tested
strength
($\alpha=0.25$) reduces the per-channel degradation to \ovheadsmooth{}
against \ovheadplain{}, short of uncalibrated group:128
(\ovheadgone{}), and no strength recovers the vocoder of Supertonic.

Table~\ref{tab:calib} applies both methods to the vocoder of F5-TTS and
the depth transformer of Kyutai, the most sensitive component of each.
Per-layer GPTQ recovers both. At group:128 the vocoder
degrades by \fvocgptqg{} and the depth transformer by \kydepgptqg{}
$\Delta$UTMOS with WER at fp, against \fvocplaing{} and \kydepplaing{}
for RTN, and the per-channel results lie between these values. Scaling is less effective in every cell of Table~\ref{tab:calib}.
Calibration is therefore not a refinement of
survivable configurations but the step that recovers the sensitive
component.

\begin{table}[t]
  \centering
  \small
  \renewcommand{\arraystretch}{0.9}
  \setlength{\tabcolsep}{4pt}
  \caption{Calibrated PTQ on the most sensitive component of each model. Paired $\Delta$UTMOS against the model's own fp baseline under RTN, activation-aware weight scaling ($\alpha=0.5$), and per-layer GPTQ at per-channel and group:128 scales. WER stays at its fp level (\ffpwer{} for F5-TTS, \kyfpwer{} for Kyutai, \vxfpwer{} for VoxCPM) in every cell except the four marked ones, whose WER is $^{a}$\kydepplainwer{}, $^{b}$1.01, $^{c}$\kydepsmoothwer{}, and $^{d}$\vxditwer{}.}
  \label{tab:calib}
  \begin{tabular}{@{}llccc@{}}
    \toprule
    Component & Scales & RTN & Scaling & GPTQ \\
    \midrule
    F5-TTS vocoder & per-channel & \mbox{-0.94} & \mbox{-0.61} & \mbox{-0.15} \\
     & group:128 & \mbox{-0.48} & \mbox{-0.22} & \mbox{-0.07} \\
    \midrule
    Kyutai depth transformer & per-channel & \mbox{-2.99}$^{a}$ & \mbox{-1.82}$^{c}$ & \mbox{-0.29} \\
     & group:128 & \mbox{-2.98}$^{b}$ & \mbox{-0.20} & \mbox{-0.08} \\
    \midrule
    VoxCPM local DiT & per-channel & \mbox{-2.73}$^{d}$ & -- & \mbox{-0.71} \\
     & group:128 & \mbox{-1.26} & -- & \mbox{-0.41} \\
    \bottomrule
  \end{tabular}
\end{table}

\subsection{Activation quantization}
\label{ssec:activations}

At the selected weight configurations, adding per-channel A8 changes the mean UTMOS
estimates by small amounts, with WER within \actbestwerbound{} of fp. W8
with A8 yields \aeightweightst{}, \aeightweightkk{}, and
\aeightweightov{} $\Delta$UTMOS on Supertonic, Kokoro, and OmniVoice,
and the W4 configurations move from \actbeststrecipe{} to \actbeststrecipea{},
from \actbestkkrecipe{} to \actbestkkrecipea{}, and from
\actbestovrecipe{} to \actbestovrecipea{}, respectively. Per-tensor A8
on the vocoder of Supertonic alone, by contrast, yields \actptvoc{} at
WER \actptvocwer{}, which per-channel and per-token scales reduce to
\actpcvoc{} and \acttokvoc{}. On OmniVoice, per-tensor A8 on the token
head alone yields \actptovproj{}, against \actptovcodec{} on the codec
decoder alone and \actptovlm{} on the LM alone, so the most sensitive
component moves from the codec decoder under weight-only PTQ to the
token head under per-tensor A8.

On the replication models, per-channel A8 on the whole model changes
UTMOS by at most \aeightpcmax{} (CSM), and adding it to their W4 configurations
costs at most \aeightrecipemax{} (Orpheus). Per-tensor A8 severely
degrades F5-TTS (\aeightptf{}), Kyutai (\aeightptky{}), CSM
(\aeightptcsm{}), and Zonos (\aeightptzn{}), and degrades the remaining four models by
\aeightptrestmin{} to \aeightptrestmax{}.

\subsection{Deployment measurements}
\label{ssec:deployment}

On the Mac mini M4 Pro, the ONNX Runtime dynamic int8 path of
Supertonic, which uses per-tensor activation scales, raises the WER from
\stfpwer{} to \stdynwer{} and the Korean CER from \kofpcer{} to
\kodyncer{} at \rtfslowfull{} the fp32 latency (Table~\ref{tab:system}).
An x86 control with int8 restricted to the MatMul operators and the
convolutions left in fp32 (MatMul-only) is quality-preserving (UTMOS
\dynmmutmos{}, WER \dynmmwer{}), so the operator coverage of the quantized graph
determines the outcome. Vocoder-excluded dynamic int8 keeps WER and CER near fp
(\stnovocutmos{} $\Delta$UTMOS) at \rtfslownovoc{} the latency and
\energycost{} more energy per audio-second (Table~\ref{tab:system}).

\begin{table}[t]
  \centering
  \caption{System metrics on the quiet Mac mini M4 Pro over 5 repeats
  that agree within 2\%. RTF and its multiple of the fp row of the same model, peak
  RSS, power P, and the marginal energy E of the whole 20-sentence process
  in J per audio-second (Sec.~\ref{sec:setup}, not measured for the int4
  row); dyn8 is dynamic int8.}
  \label{tab:system}
  \small
  \renewcommand{\arraystretch}{0.9}
  \setlength{\tabcolsep}{1.6pt}
  \begin{tabular}{@{}lrrrrr@{}}
    \toprule
    Condition & RTF & $\times$fp & RSS (MB) & P (W) & E \\
    \midrule
    Supertonic fp32 (NFE 8) & 0.148 & -- & 607 & 19.1 & 3.25 \\
    \hspace{1.6mm}dyn8 full (real int8) & 0.310 & 2.09$\times$ & 329 & 9.7 & 3.21 \\
    \hspace{1.6mm}dyn8, vocoder excl. & 0.218 & 1.47$\times$ & 436 & 19.1 & 4.64 \\
    \hspace{1.6mm}int4 W, vocoder excl. & 0.089 & 0.60$\times$ & 360 & -- & -- \\
    \midrule
    Kokoro fp32 & 0.069 & -- & 2712 & 7.5 & 0.99 \\
    \hspace{1.6mm}dyn8 (real int8) & 0.077 & 1.12$\times$ & 2722 & 8.9 & 1.25 \\
    \bottomrule
  \end{tabular}
\end{table}

The effects of compiled GPU int8 differ across models and devices. Weight-only
int8 on the LM of Orpheus runs
\orpintlat{} faster than compiled bf16 at \orpintenergy{} less energy on
an L4, whereas the sign of the energy change of the W8A8 path of
OmniVoice depends on the device. At one fully controlled operating point
(NFE 8, compiled, RTX PRO 6000, 200 sentences), compiled fp16,
unquantized bf16, and W8A8 score within 0.02 UTMOS of one another, while
W8A8 requires \ovmatchedlat{} the latency and \ovmatchedenergy{} the
energy of compiled fp16, against \ovbfmatchedlat{} and
\ovbfmatchedenergy{} for bf16.

Real 4-bit weight kernels reproduce the simulated ordering. Supertonic
under \texttt{MatMulNBits} degrades by \stnbitsall{} $\Delta$UTMOS with
the vocoder included and by \stnbitsnovoc{} with the vocoder excluded
(WER \stnbitsnovocwer{} against \stfpwer{} at fp, Korean CER
unchanged), against \stsimnovocg{} for the simulated group:128
configuration. On the Mac mini, this vocoder-excluded 4-bit path runs at
\stnbitsrtfratio{} the fp32 latency with \stnbitsrsssave{} lower peak
RSS (Table~\ref{tab:system}), whereas the int8 path is slower than fp32. torchao
int4 weight-only quantization of the LM yields \ovintfour{} on OmniVoice
(simulated \ovsimlmg{}, Korean CER \ovintfourkocer{}
\ovintfourkocerci{}), \orintfour{} on Orpheus, \cbintfour{} on
Chatterbox, and \vxintfour{} on VoxCPM, and it reproduces the severe
degradation of the depth transformer of Kyutai (\kyintfour{}) and the
degradation of CSM (\csmintfour{}). On the RTX PRO 6000, int4 on OmniVoice at NFE 8 requires
\ovintfourlat{} the latency and \ovintfourenergy{} the energy of compiled
fp16, whereas on the larger LM of Orpheus it runs at \orintfourlat{} the
latency and \orintfourenergy{} the energy of bf16.

\subsection{Staged procedure and blind application}
\label{ssec:procedure}
\label{ssec:blind}

The results suggest this order of evaluation. Apply separate
intelligibility and naturalness criteria against matched fp baselines,
evaluate weight-scale granularity at fixed precision
(Sec.~\ref{ssec:granularity}), and where the degradation persists,
identify the sensitive component by ablation and protect it with higher
precision or calibrated PTQ (Secs.~\ref{ssec:component}
and~\ref{ssec:calibrated}). Then evaluate activation scaling on its own
(Sec.~\ref{ssec:activations}), the selected settings in combination, and
finally quality and system cost on the target runtime
(Sec.~\ref{ssec:deployment}). Chatterbox and VoxCPM were quantized after
this procedure was fixed, with the sensitive component predicted from the
sensitivity map and recorded before any run (the vocoder for
Chatterbox, the VAE decoder or the LM for VoxCPM). Both predictions were incorrect. The procedure found the flow-matching
decoder in each case, which neither the model class nor the parameter share
predicted (Fig.~\ref{fig:map}). On Chatterbox, the T3 backbone alone yields
\cblm{} $\Delta$UTMOS at W4, the flow decoder alone \cbflow{}, and
everything except the flow decoder \cbrest{}, which equals the sum of the
individual losses. Group:128 scaling reduces the flow-decoder loss to
\cbflowg{}, keeping S3Gen at W8 with T3 at W4 per-channel yields
\cbrecipe{} at fp-level WER, and real int4 on T3 yields \cbintfour{}. On VoxCPM, the LM, the residual LM, the encoder, and the projections
change UTMOS by at most \vxminormax{} in magnitude, whereas
the local DiT alone degrades by \vxdit{}, the VAE decoder by \vxvae{},
and everything except the DiT by \vxrest{}, which exceeds the sum of the
individual losses. Group:128 scaling reduces the VAE decoder loss to \vxvaeg{} but only
halves the DiT loss (\vxditg{}). Per-layer GPTQ
improves the DiT to \vxditgptqg{} at group:128 without reaching the
quality-preserving band (Table~\ref{tab:calib}), and the all-W4 configuration
with both components calibrated yields \vxallcalib{}, whereas keeping both at W8 with the remainder at W4
group:128 yields \vxrecipeg{} at fp-level WER. Real int4 on the LM
yields \vxintfour{}.

\subsection{Quality-preserving configurations}
\label{ssec:quality}

Nine configurations meet both conditions of Sec.~\ref{sec:setup} (five
rows of Table~\ref{tab:crit}, the W8 results of
Sec.~\ref{ssec:granularity}, and the MatMul-only control of
Sec.~\ref{ssec:deployment}), three of them at 4 bits, namely Kokoro W4 group:128 and the Chatterbox configuration and real
int4 path. The other rows
of Table~\ref{tab:crit} fail the UTMOS condition by 0.01 to 0.06 at
fp-level WER, except the Orpheus int4 path, which fails the WER
condition by 0.002.

\begin{table}[t]
  \centering
  \small
  \renewcommand{\arraystretch}{0.9}
  \setlength{\tabcolsep}{2.5pt}
  \caption{The quality-preserving criterion applied. The paired 95\% interval of $\Delta$UTMOS (CI) must lie within $[-0.05, 0.05]$, and the upper endpoint of the paired $\Delta$WER interval (WER$^{+}$) must be at most $0.01$, unrounded, against the fp baseline of each configuration. An asterisk marks a violated condition.}
  \label{tab:crit}
  \begin{tabular}{@{}lcc@{}}
    \toprule
    Configuration & CI & WER$^{+}$ \\
    \midrule
    Kokoro real int8 (PyTorch) & [\mbox{-0.000}, 0.003] & 0.001 \\
    OmniVoice W8A8, compiled, NFE 8 & [\mbox{-0.038}, 0.048] & 0.002 \\
    Kokoro W4 group:128 & [\mbox{-0.039}, \mbox{-0.035}] & 0.002 \\
    Supertonic int8, no vocoder & [\mbox{-0.056}, \mbox{-0.018}]$^*$ & 0.005 \\
    Chatterbox T3 W4, S3Gen W8 & [\mbox{-0.032}, 0.001] & 0.005 \\
    Chatterbox real int4 on T3 & [\mbox{-0.012}, 0.020] & 0.003 \\
    Orpheus real int4 on the LM & [\mbox{-0.022}, 0.034] & 0.012$^*$ \\
    Kyutai depth transf. W4 g128 GPTQ & [\mbox{-0.111}, \mbox{-0.058}]$^*$ & 0.006 \\
    F5-TTS vocoder W4 g128 GPTQ & [\mbox{-0.088}, \mbox{-0.046}]$^*$ & 0.004 \\
    Supertonic int4 W, no vocoder & [\mbox{-0.094}, \mbox{-0.042}]$^*$ & 0.009 \\
    \bottomrule
  \end{tabular}
\end{table}

\section{Limitations}
\label{sec:limitations}

The two blind applications (Sec.~\ref{ssec:blind}) test whether the
procedure finds the sensitive component, not the benefit of its ordering,
and both held-out models share the codec-token or latent-patch design of the sensitivity map. UTMOS is trained on English. NISQA-TTS
shows positive rank correlations with UTMOS across the conditions of each
English model (Spearman \nisqarhomin{} to \nisqarhomax{} on ten of the eleven English
baselines and \nisqarhosttwo{} on StyleTTS~2), but the
Korean evidence in the text remains CER, and the reported UTMOS improvements concern predicted naturalness
only.\footnote{An informal listening test (\listenraters{} raters,
\listenclips{} clips, clip-level Spearman $\rho$ = \listenrho{} with
UTMOS) agrees with UTMOS on severe degradation (condition MOS at most
2.0) but not consistently on moderate differences. Listeners rated the
VoxCPM mixed-precision configuration above its DiT W4 group:128 GPTQ variant despite the lower
UTMOS, so the test validates neither moderate UTMOS differences nor
configuration rankings.} The bootstrap intervals quantify sentence sampling
and do not correct for the selection of the best strength or granularity
on the test sentences. Across-seed standard deviations of the paired $\Delta$UTMOS
(\nseedcond{} conditions, \nseedmodels{} models) stay below \seedsdmax{}.

\section{Conclusion}
\label{sec:conclusion}

Across three core, eight replication, and two held-out TTS models, the
same bit width yields different outcomes because the sensitive component
is model-specific and was not reliably predicted from the model class
alone. A staged ablation procedure with matched fp baselines identified
that component in two blind tests, and higher precision or per-layer
GPTQ can restore it to within 0.1 UTMOS of fp. The waveform decoder or codec is the first component to test, and
per-tensor scaling can cause severe degradation even with 8-bit weights
or activations.

\bibliographystyle{IEEEbib}
\bibliography{refs}

\end{document}